\documentclass[12pt]{iopart}

\newcommand{\Om}{\ensuremath{\Omega}}

\newcommand{\Lam}{\ensuremath{\Lambda^0}}
\newcommand{\kos}{\ensuremath{K^0_S}}
\newcommand{\pit}{\ensuremath{p_{\perp}}}
\newcommand{\dphi}{\ensuremath{\Delta\phi}}
\newcommand{\deta}{\ensuremath{\Delta\eta}}
\newcommand{\vtwo}{\ensuremath{v_{2}}}

\usepackage{iopams}
\usepackage{graphicx}

\begin{document}

\title[Strange Particle Correlations in STAR]{High p$_{T}$ correlations with strange particles in STAR}

\author{B I Abelev (for the STAR Collaboration)}

\address{University of Illinois at Chicago,
Chicago, Illinois, 60607}
\ead{betya@uic.edu}
\begin{abstract}
We present the highlights of the current identified strange particles \dphi\ and \deta\ correlations analyses, including system-size and trigger-\pit\ of the jet and ridge, jet, ridge and away-side meson/baryon ratios, and the current state of the multi-strange baryon analysis.  We see clear azimuthal peaks of comparable strength for all strange baryons and \kos\ mesons.  We see no observable species dependence on the same-side jet or ridge yields as a function of \pit.\ However, while the away side and the ridge have \Lam\ to \kos\ ratio similar to that of the bulk, the jet-only ratio is similar to that in $p+p$. The implications of these findings on current in-medium jet theoretical explanations are discussed.
\end{abstract}


\section{Introduction}
When discussing particle production in Relativistic Heavy Ion collisions, it is generally understood that low \pit\ ($<$ 1.5 GeV/c) particles are produced thermally, while the high end of the \pit\ ($>$ 10 GeV/c) particle spectra are dominated by fragmentation.  To trace the dwindling of a particle production mechanism dominance and the onset of another, we examine the intermediate regime (2-5 GeV/c).  At RHIC, strange particles make excellent probes in this \pit\ region.  All strange particles observed are created during the course of the collision. In addition, topological reconstruction of strange baryons and $K^0_S$ mesons allows access to identified particles with \pit\ up to at least 7 GeV/c \cite{Matt}.  The STAR detector, due to the full azimuthal and three units of pseudo-rapidity coverage of its large acceptance Time Projection Chamber (TPC), is particularly well-suited for strangeness reconstruction.  Particle spectra studies show an excess of baryons at mid-\pit\ in Au+Au collisions w.r.t. $p+p$ collisions, with a peak at ~3 GeV/c \cite{Matt}.  This indicates the transition region from thermal to fragmentation regime for singly strange particles.

A way to study fragmentation is to study the resulting particle jets. In a high multiplicity environment, such as central Au+Au collisions, full jet reconstruction is very difficult.  However, we can still access jet production via a statistical study of jet products.  This is possible due to the inherent azimuthal symmetry of a di-jet and the characteristic cone-shape of the jet itself.  This method has proved to be very successful for unidentified charged particles \cite{Mike}. Reconstruction of strange particles using their decay topology yields very pure samples of particles with known \pit.\  This allows us to calculate particle's azimuthal angle at the event vertex and then treat the reconstructed strange hadron in the same manner as unidentified one.

 Recently, an elongation in \deta\ of the same-side peak, known as the ``ridge," has been discovered in central and mid-central Au+Au collisions.  The origin of the ridge is still under investigation.  Current theories include ``phantom jets" \cite{Rudy2}, medium-induced broadening \cite{Armesto}, gluon propagation through color field instabilities \cite{Bass}, and others.  In this proceedings we present the measurement of the ridge both as a system-size dependent variable, and a variable that depends on the \pit\ of the leading hadron.

\section{Analysis methods and data sets}
In these analyses we use high \pit\ correlation functions where one of the correlated particles is a strange particle, identified through its decay topology. A correlation consists of a trigger and an associated particle.  The trigger particle is assumed to be the leading jet particle, thus it is always the particle with the higher \pit.\  The data presented come from $d$+Au, Cu+Cu and Au+Au data sets, all taken at top RHIC energy.  In the case of a strange particle-triggered correlation we first build a high purity invariant mass sample of the trigger particle in a given \pit\ range.  Only those particles with reconstructed mass less than two standard deviations away from the centre of the peaks are used in the analysis.  The event is then scanned for an associated track, which can be any particle above an associated \pit\ threshold and below that of the trigger.  We also ensure the associated particle is not a trigger decay product and comes from the collision vertex. If a suitable pair of particles is found, we calculate \deta\ and \dphi\ between them. After all available pairs are exhausted, we normalize the correlation by the number of triggers.  In the case of strange associated particles we first look for a trigger hadron, which can be any charged track originated at the primary event vertex with sufficient \pit.\  We then look for a suitable strange correlation partner in the same event, compute \deta\ and \dphi,\ and normalize per trigger as before.

An important component of this study is identifying and subtracting the underlying background. In $d$+Au data the background consists of random pairs of uncorrelated particles, and is flat.  We determine its height by fitting the correlation using a two Gaussian curves and a flat line ansatz.   In Cu+Cu and Au+Au collisions the background is modulated by the second harmonic of the underlying flow, which is sinusoidal and symmetric about the azimuth.  The background can then be expressed as

\begin{equation}\label{eq:AuAuBG}
    f(\Delta\phi) = B(1+2v_2^{trig}v_2^{assoc}\cos{2\Delta\phi})
\end{equation}

In this equation, $B$ is the height of the uncorrelated background, while $v_2^{trig}$ and $v_2^{assoc}$ stand for the elliptic flow coefficient of the trigger and associated particle respectively.  Additionally, all correlations are corrected for the efficiency of the associated particle and for detector acceptance. The efficiency correction is done using particles embedded in real events.  We corrected for the detector acceptance using the event-mixing technique.  The functions and resultant yields in Figures \ref{fig:la_xi_om} and \ref{fig:auau_yields} were corrected for azimuthal acceptance only, while the data presented in all other figures were corrected for both azimuthal and pseudo-rapidity acceptance.

In all correlations, including those where both the associated and the trigger hadrons are unidentified charged tracks, we observe a depression at $\deta\simeq\dphi\simeq0,$ known as the ``dip." The dip is seen to be comprised of five smaller dips.  Four of these, symmetrical about $\deta\simeq\dphi\simeq0,$ are thought to be due to merging of a \Lam\ or a \kos\ daughter and associated tracks while crossing in the TPC.  Each dip is due to a particular helicity combination, causing the loss of available trigger-associate pairs in a specific region of the \deta-\dphi\ space.  The dip in the center is due to genuine track merging unrelated to helicity combination of the pair \cite{Marek}.  This is approximately a 10\% effect for h-h, and about 15\% for $\Lambda^0$-h and $K^0_S$-h correlations in Au+Au.  The effect significance is yet unknown for multi-strange correlations or those in Cu+Cu. A ``mirror image" method was developed to correct for the helicity-dependent dips, and used effectively for Au+Au \Lam\ and \kos\ data \cite{Marek}.

\section{Results}
\begin{figure}[t]
\begin{minipage}[t]{8cm}
\centering
  \includegraphics[width=0.95\textwidth]{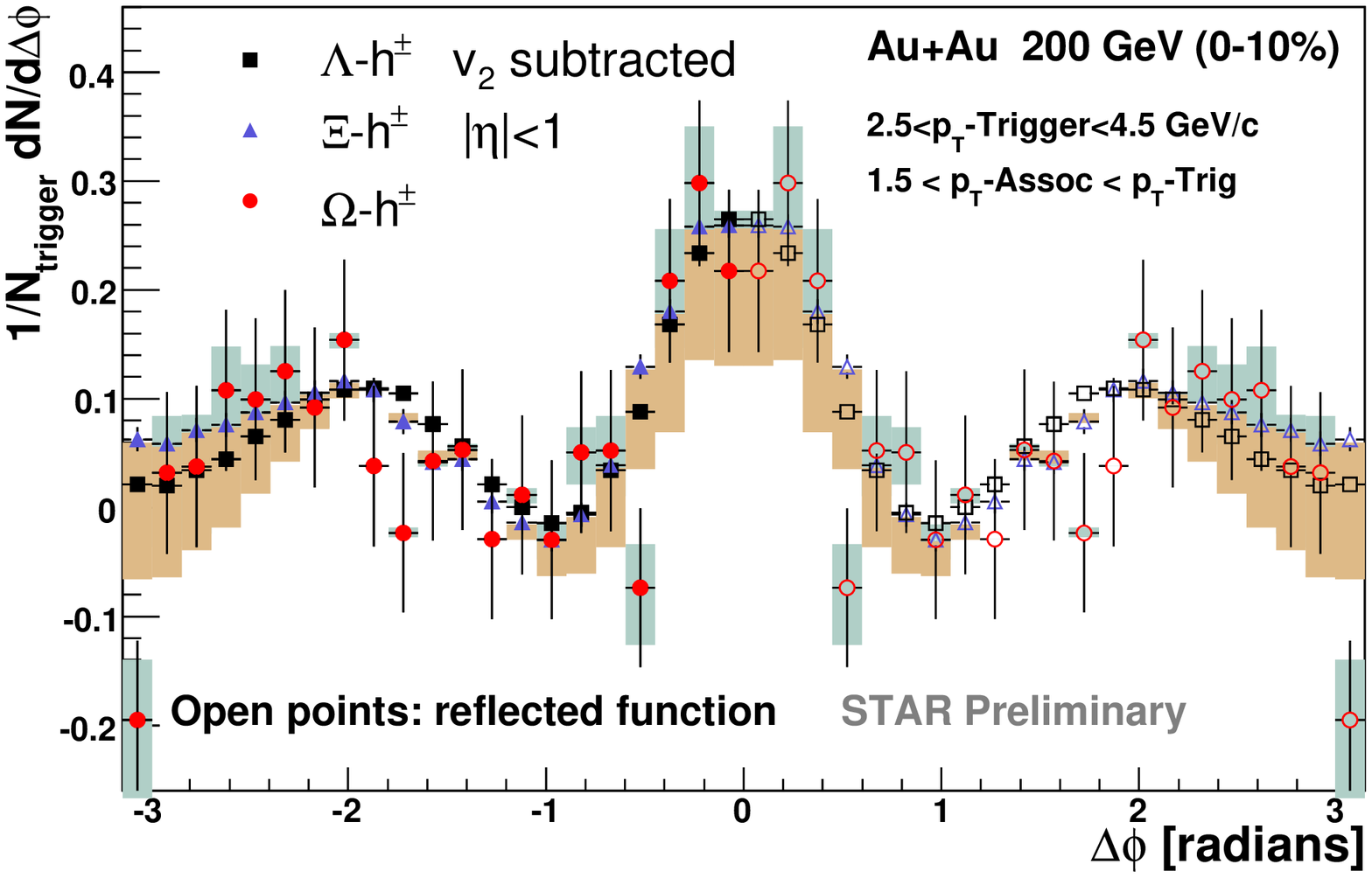}\\
  \caption{Azimuthal correlations constructed with $\Omega$, $\Xi$, and \Lam\ triggers in Au+Au $\sqrt{s_{NN}}=200$ GeV data, normalized per trigger.  $p_{\perp}$-trigger is between 2.5 and 4.5 GeV/c, and $p_{\perp}$-associated above 1.5 GeV/c. Shaded bands represent uncertainties due to \vtwo\ determination.}\label{fig:la_xi_om}
\end{minipage}
\hfill
\begin{minipage}[t]{8cm}
\centering
  \includegraphics[width=0.95\textwidth]{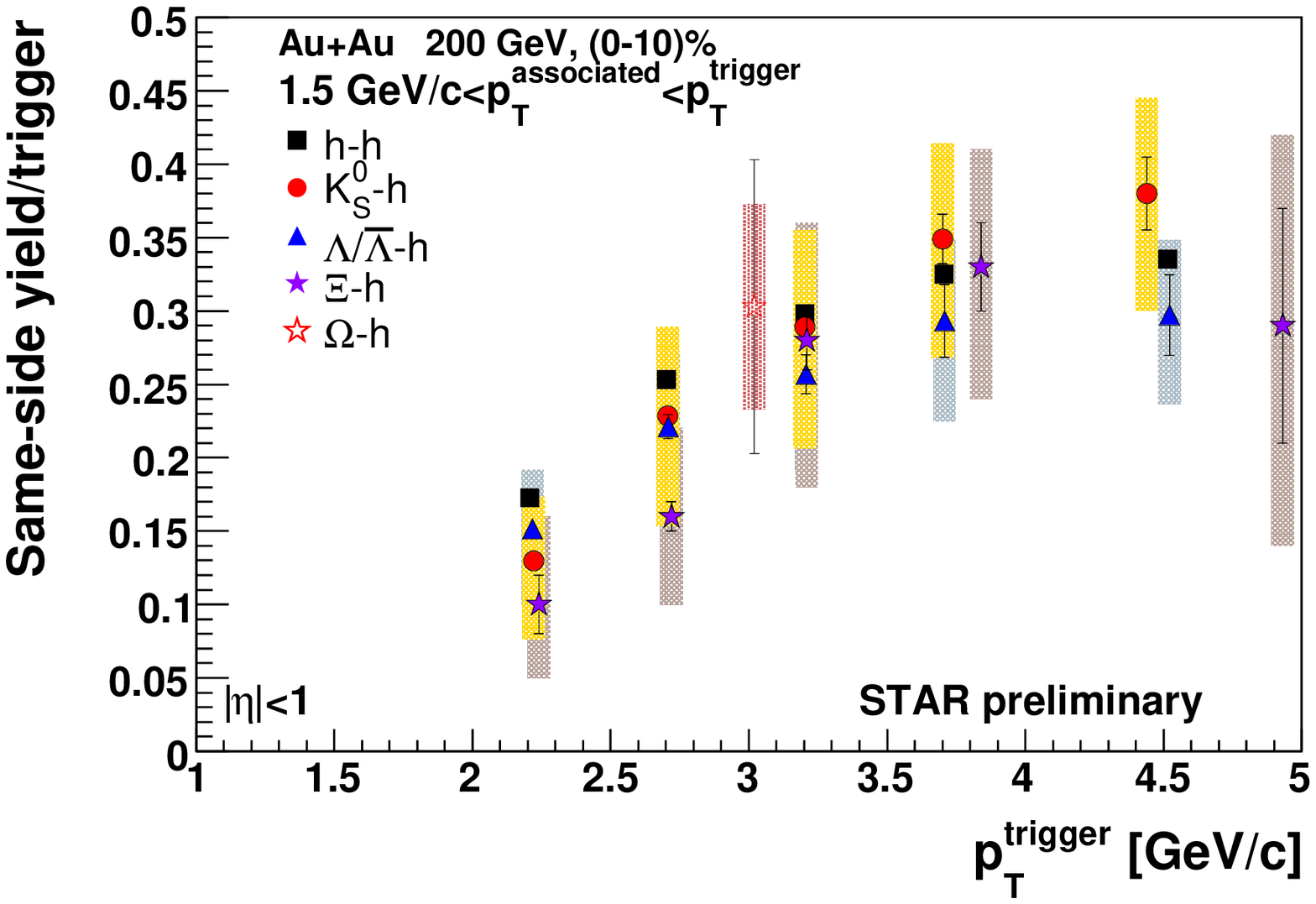}\\
  \caption{Jet+ridge combined yields of h-h and strange particle correlations in 0-10\% central Au+Au collisions. The shaded bands represent systematic uncertainty due to \vtwo\ determination.}\label{fig:auau_yields}
\end{minipage}
\end{figure}

In 0-10\% central Au+Au collisions, we observe the emergence of jet-like same-side \dphi\ correlation structures at mid-$p_{\perp}$ (2.5 $<p_{\perp}$-trigger $<$ 4.5 GeV/c) for all strange baryon species, as illustrated in Figure \ref{fig:la_xi_om}.  Although the statistics for the \Om\ correlation are quite poor, it is evident that the strength of the same-side correlation peak is very similar to those of $\Xi$ and $\Lambda$ baryons triggered correlations. Moreover, we can extract same-side yields as a function of trigger $p_{\perp}$, as shown in Figure \ref{fig:auau_yields}.  The yields are consistent across all species, including the \Om\ baryons.  This is an important result, as initial predictions based on the recombination model forecasted no same-side \Om\ jet in azimuth \cite{Rudy}.  Since the finding of this \Om\ same-side peak, the prediction was revised to include the ridge phenomenon \cite{Rudy2}. According to the new predictions, the \Om\ peak is entirely due to ``phantom jets", which arise when non-strange partons fragment in a medium.  Because there are insufficient statistics for a two-dimensional study of \Om\ baryon correlation, we can examine the ridge using particles with fewer $s$-quarks.   As statistics are also insufficient to do a ridge subtraction for $\Xi$ baryon correlation (for method see, for example, \cite{Christine}), we examine the ridge only in singly-strange triggered correlations. 

\begin{figure}[t]
\begin{minipage}[t]{8cm}
\centering
  \includegraphics[width=0.8\textwidth]{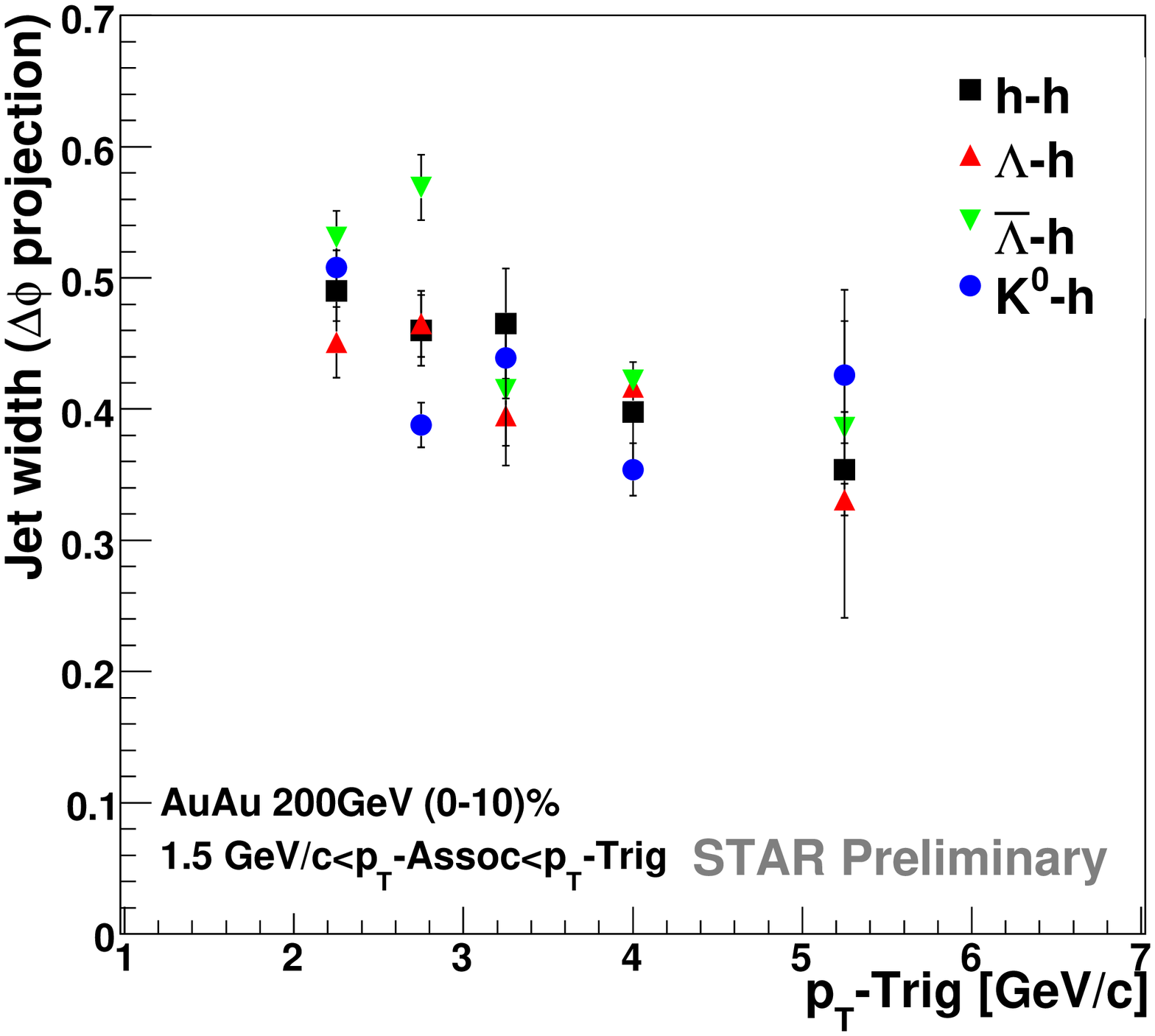}\\
  \caption{The width of the same-side jet only peak in $\Delta\phi$ as a function of trigger $p_{\perp}$.}\label{fig:dphiwidth}
\end{minipage}
\hfill
\begin{minipage}[t]{8cm}
\centering
  \includegraphics[width=0.8\textwidth]{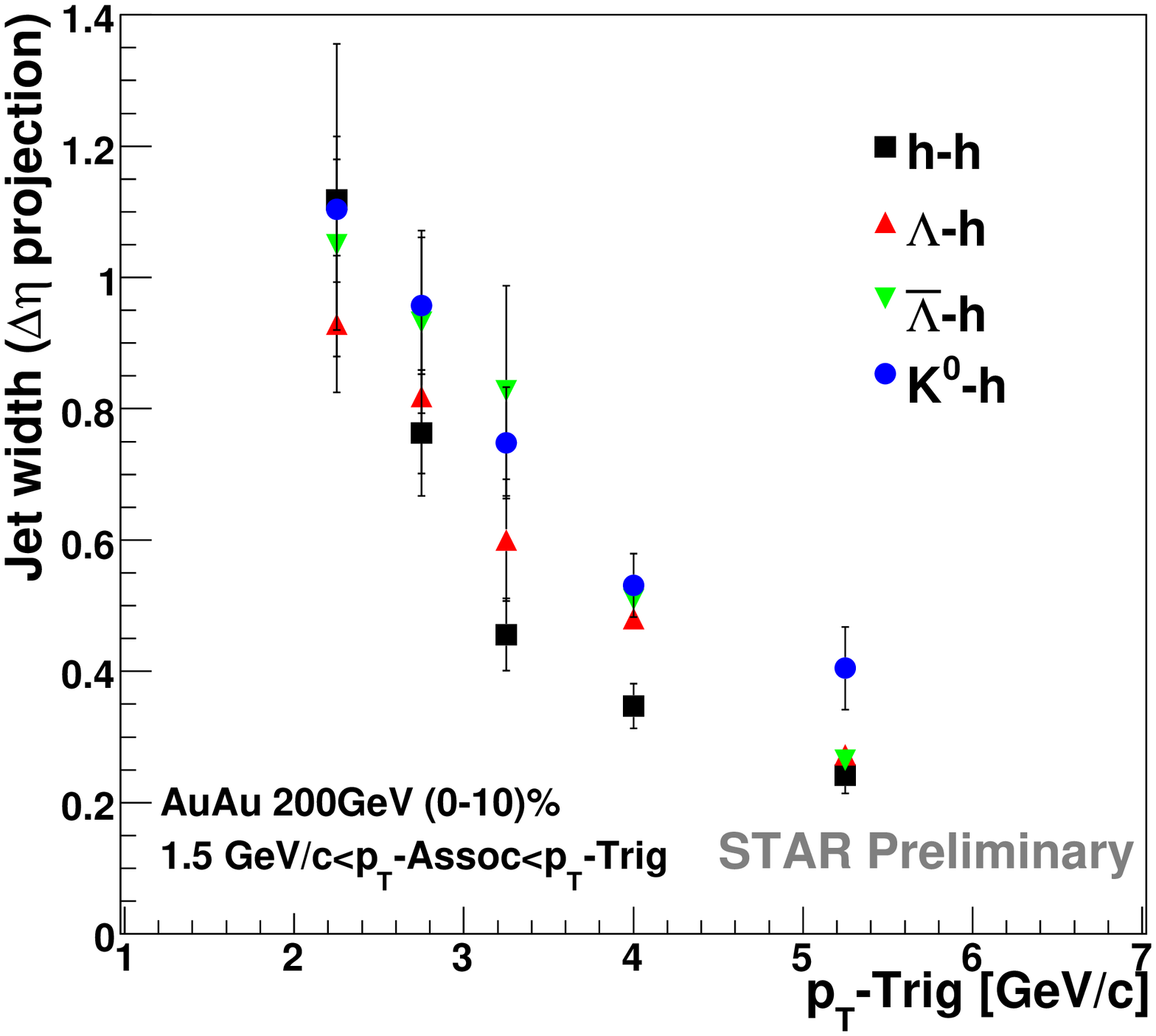}\\
  \caption{The width of the same-side jet only peak in $\Delta\eta$ as a function of trigger $p_{\perp}$.}\label{fig:detawidth}
\end{minipage}
\end{figure}

For both \Lam\ and \kos\ correlations we can extract the width of the jet cone in both \deta\ and \dphi\, as seen in Figures \ref{fig:dphiwidth} and \ref{fig:detawidth}.  The \deta\ projection of the jet cone is taken for $|\Delta\phi|<1$, while the width of the jet peak in \dphi\ is calculated using $\Delta\phi(|\Delta\eta|<1)-\Delta\phi(1<|\Delta\eta|<2$).  Again, we see no difference between the species. Yet we observe a narrowing of the jet-only peak width in \deta\, while the \dphi\ peak width stays constant.   We attribute this to a growing strength of the jet with increasing \pit,\ and to the saturation of the ridge component of the jet, as was shown previously \cite{Jana}.

\begin{figure}[t]
\begin{minipage}[t]{8cm}
\centering
  \includegraphics[width=0.75\textwidth]{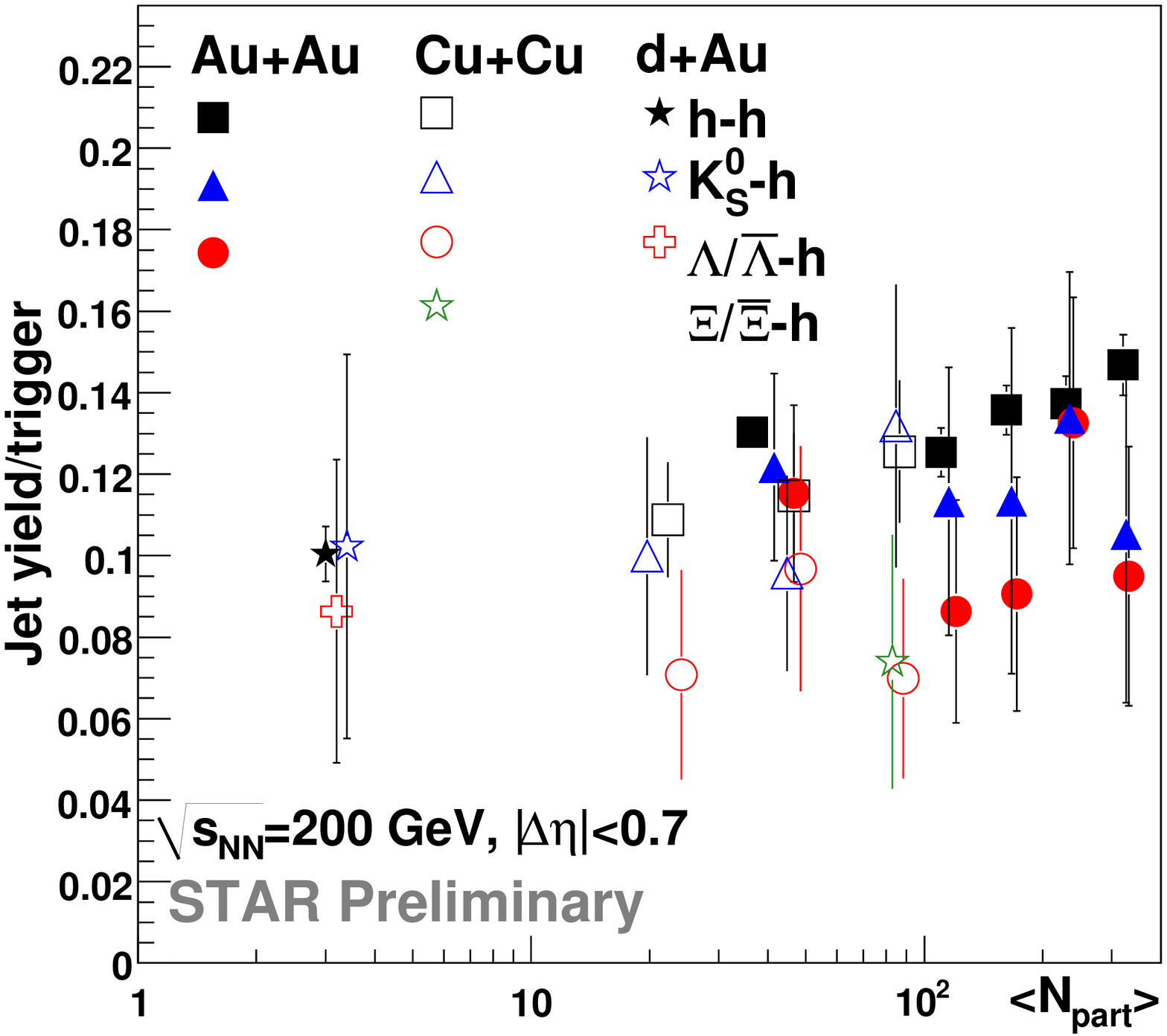}\\
  \caption{Dependence of jet yields on N$_{part}$ in $d$+Au, Cu+Cu, and Au+Au collisions for charged hadron, \kos,\ $\Lambda^0$, $\Xi$, and $\Omega$ triggered correlations.}\label{fig:jetYields}
\end{minipage}
\hfill
\begin{minipage}[t]{8cm}
\centering
  \includegraphics[width=0.75\textwidth]{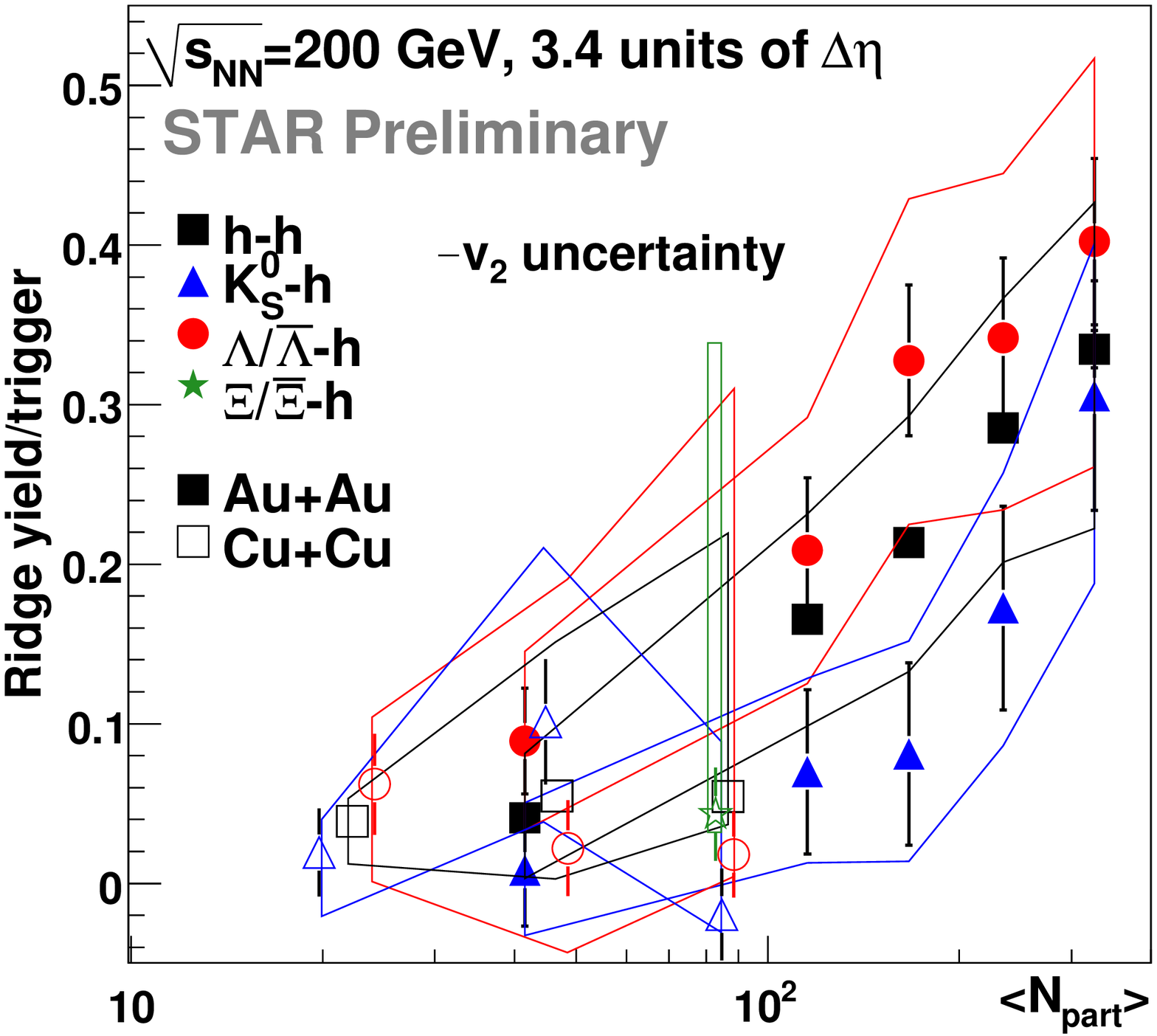}\\
  \caption{Dependence of ridge yields on N$_{part}$ in $d$+Au, Cu+Cu, and Au+Au collisions. Bands represent the systematic uncertainty due to \vtwo\ determination.  The ridge is measured over 2 units in $\Delta\eta$ and scaled to 3.5 units.}\label{fig:ridgeYields}
  \end{minipage}
\end{figure}

Decomposing the same side into jet and ridge components in Cu+Cu, as shown in Figures \ref{fig:ridgeYields} and \ref{fig:jetYields}, we observe that the result fits well into the picture previously seen in Au+Au \cite{Jana}.  The jet (Figure \ref{fig:jetYields}, obtained from $|\Delta\eta|<$0.7), stays constant as a function of nucleons participating in the collision ($N_{part}$), and is consistent with the vacuum ($d$+Au) result.  The ridge (Figure \ref{fig:ridgeYields}, measured in $0.75<|\Delta\eta|<1.75$ and scaled to 3.5 units in $\Delta\eta$), however, is consistent with zero in Cu+Cu and rises monotonically as a function of $N_{part}$.

\begin{figure}[t]
\begin{minipage}[t]{8cm}
\centering
  \includegraphics[width=0.9\textwidth]{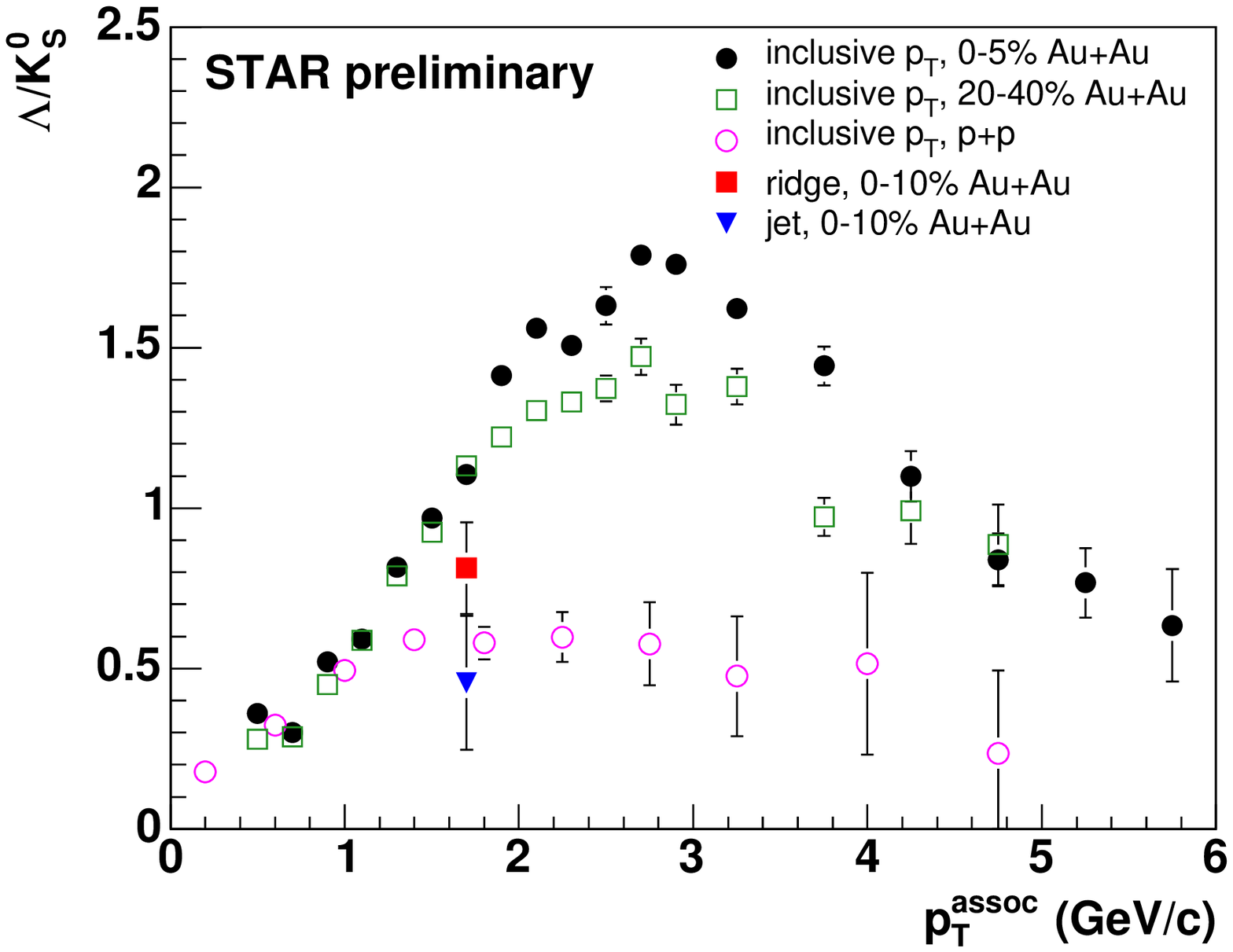}\\
  \caption{\Lam/\kos\ ratios in jet and ridge (closed square and triangles, respectively). }\label{fig:ridgejetratio}
\end{minipage}
\hfill
\begin{minipage}[t]{8cm}
\centering
  \includegraphics[width=0.9\textwidth]{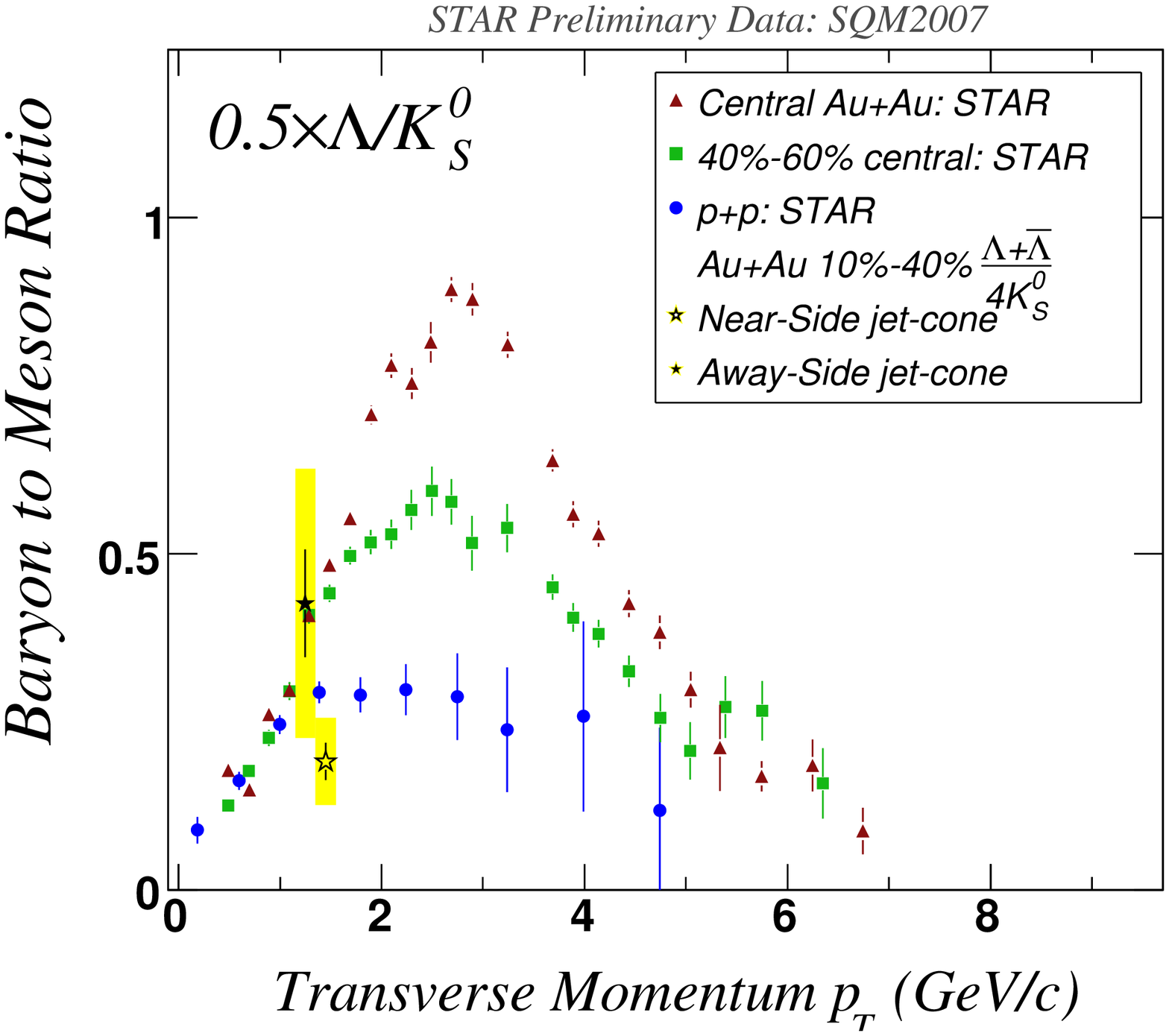}\\
  \caption{\Lam/\kos\ ratios in the same (open stars) and away (closed stars) side of a correlation}\label{fig:sameawayratio}
\end{minipage}
\end{figure}

 We now compare the ridge and jet spectra to the spectra of the bulk. We also measure the spectrum of the away side.  From previous studies we know that the temperature of the ridge in Au+Au collisions is similar to that of the medium \cite{Jana}.  We also see this in Cu+Cu \cite{Christine}.  In order to understand the behavior of strange particles in the ridge, we use them as associated particles. The result is shown in Figures \ref{fig:ridgejetratio} and \ref{fig:sameawayratio}.  We compare the total jet, ridge, and the away side yields to the strange baryon to meson ratio measured in the inclusive data.  In Figure \ref{fig:ridgejetratio} the $p+p$ data is depicted as open circles, while $Au+Au$ data is shown in closed circles (most central collisions) and open squares (mid-central data).  At \pit\ as low as 2 GeV/c we observe that the strange associated particle baryon to meson ratio found in the jet closely matches that observed in $p+p$ collisions (with no medium), while the ridge lies much closer to the ratio found in Au+Au collisions, where we expect a produced medium.  In Figure \ref{fig:sameawayratio} we compare the ratio in the same and away side of the correlation to the ratio in the inclusive data. Here the Au+Au data are depicted as closed squares and triangles, while the $p+p$ data are shown as closed circles.  Yet again, we observe the ratio of the same-side yields to resemble that of the bulk. The ratio of the away-side yields matches that seen in the medium-free $p+p$ environment, suggesting that it is a result of a fragmentation process \cite{Jiaxu}.  Thus, it appears that the models \cite{Rudy2,Bass}, which explain the existence of the ridge as a medium effect, do so accurately.  However, at the moment we have no means to distinguish between them.

\section{Conclusions and outlook}
In this article we have presented the highlights of the current identified strange particles correlations research. We have seen no appreciable species dependence on the same-side jet or ridge yields as a function of \pit.\  However, we observe clear differences in the composition of the same and away side peaks, as well as a difference in the composition of the jet component and the accompanying ridge.  The ridge and the away side are composed of the same baryon-meson mixture as the bulk created in the medium-producing (Au+Au, Cu+Cu) collisions, while the jet itself is consistent with particle ratios in fragmentation-dominated collisions, such as $p+p$. Taking into account the near-disappearance of the ridge in the peripheral and mid-central Cu+Cu collisions, this suggests that the ridge is a property of in-medium parton fragmentation, dependent on medium density and limited to a specific \pit\ range.

Up to now the analyses of jet and ridge composition were performed using only singly strange hadrons.  Thus it is difficult to disentangle the effects of the medium on the $u$ and $d$ quarks from those on the $s$-quark.  Identifying the ridge and the jet component in the multi-strange correlations will further our understanding of the processes responsible for production of the $s$-quark.  Thus, the two-dimensional analyses of the $\Xi$-h and $\Omega$-h correlations are a priority.  With the high statistic Au+Au data taken by the STAR experiment this year, we expect to further the resolution of this puzzle. In addition, the determination of the high \pit\ tail of \Om\ and $\Xi$ baryon spectral shapes is under way.  With better statistics, we will be able to pin-point the change from an exponential to a power-law-like behavior to deduce the on-set of fragmentation.

\end{document}